\documentclass[10pt]{article}%
\usepackage{graphicx}
\usepackage{times}
\usepackage{amssymb}
\usepackage{amsmath}
\usepackage{amsfonts}
\title{Probing Sgr A* with  theory of large masses without events
horizon}

\author{L. Verozub\\
Kharkov National University}
\date{}
\begin{document}
\maketitle

\begin{abstract} In the present paper some consequences of the assumption that in the center
of the Galaxy there is a supermassive compact object without the events horizon are
considered. The possibility of existence of such object has been argued
 earlier. It is shown, that accretion of a surrounding gas onto the object can
cause nuclear burning in a superficial layer which owing to comptonization in a
hotter layer, laying above, can manifest itself in observable IR and X spectra. The
contribution of an intrinsic magnetic moment of the object in the observable
synchrotron radiation is considered, using transfer equations, taking into account
influence of gravitation on the energy and movement of  photons.
\end{abstract}

\twocolumn
\section{Introduction}
An analysis of stars motion in the dynamic center of the Galaxy give strong evidence
for the existence of a compact object with mass about $3\cdot 10^{6}M_{\odot}$ or
more that is associated with  Sgr A* \cite{gensel1},\cite{gensel2},\cite{ghes1},
\cite{ghes2}. There are three kinds of an explanation of observable peculiarities of
the object radiation:

1 - The gas accretion onto the central object -- a supermassive black hole (BH)
\cite{melia},\cite{narayan},

2 - An ejection of the magnetized plasma from the vicinity of the Schwarzschild
radius of the BH \cite{falcke}, \cite{melia1}

3 - Explanations based on hypotheses about another nature of the central object (a
cluster of dark objects \cite{maoz}  ,  a fermion ball hypothesis \cite{viollier},
boson stars  \cite{torres}.

In the present paper we consider some consequence of the assumption that  radiation
of Sgr A* is caused by existence of a supermassive compact object without events
horizon in the Galaxy Center. Such steady configurations of the degenerated
Fermi-gas with masses $10^{2}\div10^{10}$ $M_{\odot}$ and with the radiuses $R$ less
than the Schwarzschild radius $r_{g}$ are one of the consequence of the metric-field
equations of gravitation \cite{verozub95}, \cite{verkoch01}, \cite{verozub01}. In
the theory  gravitational field of an attractive mass manifests itself as a field in
Minkowski space-time for a remote observer in an inertial frame of reference, and as
space-time curvature for the observer in a comoving (with the free falling
particles) frame of reference. Physical consequences from the gravitation equations
under consideration are very close to the ones in general relativity at the
distances from the central mass much more than $r_{g}$ . However they are completely
different at the distances nearby $r_{g}$ or less than that. The
spherically-symmetric solution of the gravitation equations have no the event
horizon and physical singularity in the center \cite{verozub91}.

Since the gravitational equations was tested by the binary pulsar PSR 1913+16
\cite{verkoch00} and stability of the supermassive configurations was studied
sufficiently rigorously \cite{verkoch01}, it is meaningful to investigate the
possibility of the existence of such objects at the Galaxy Center as an alternative
to the supermassive black hole hypothesis.

 The gravitational force of a point mass $M$ affecting a
free falling particle  of mass $m$ is given by \cite{verozub91}.
\begin{equation}
F=-m\left[  c^{2}B_{1}+(B_{2}-2B_{3})\overset{\cdot}{r}^{2}\right]  ,
\label{gravaccel1}%
\end{equation}
where
\begin{equation}
B_{1}=C^{\prime}/2A,\ B_{2}=A^{\prime}/2A,\ B_{3}=C^{\prime}/2C
\end{equation}
and%
\begin{equation}
A=f^{\prime2}/C,\ C=1-r_{g}/f,\ f=(r_{g}^{3}+r^{3})^{1/3}.
 \label{ABC}
\end{equation}
In this equation $r$ is the radial distance from the center, $r_{g}%
=2GM/c^{2},$ $M$ is the mass of the object, $G$ is the gravitational constant,
$c$ is speed of light, the prime denotes the derivative with respect to $r$.
 The force affecting the test particle in rest is%

\begin{equation}
F=-\frac{GmM}{r^{2}}\left[  1-\frac{r_{g}}{(r^{3}+r_{g}^{3})^{1/3}}\right]
\label{ForceStat}%
\end{equation}

\begin{figure}
\resizebox{\hsize}{!} {\includegraphics[]{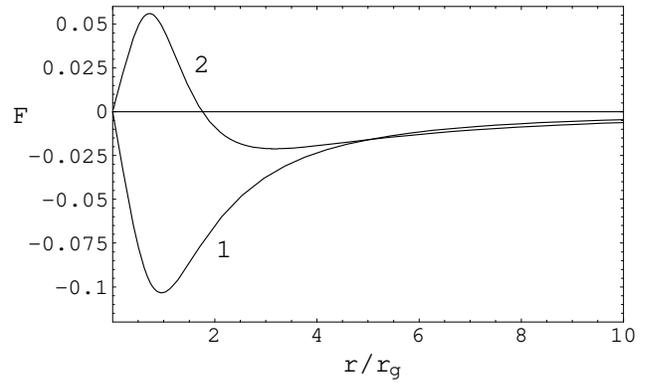}}
 \caption{The gravitational force (arbitrary units) affecting a test particle at rest (curve
1) and free falling particle (curve 2) near the point attractive mass $M$.}
 \label{GRForce}
\end{figure}

Fig. \ref{GRForce} shows the force $F$ (in arbitrary units ) affecting the test
particle at rest (curve 1) and the free falling particle (curve 2) as the function
of the distance $\overline{r}=r/r_{g}$ from the center.

It follows from the above plot that the gravitational force affecting free falling
particles decreases when $r$ approach to $r_{g}$ and changes its sign at
$\thicksim1.5$ $r_{g}$ . Although we never observed the motion of the particle at
distances close to $r_{g}$ , we can test this conclusion for very distant objects in
our Universe because for its observed mass $M_{u}$ the value of $2GM_{u}/c^{2}$ is
close to the observed radius $r_{u}$ of the Universe. At such distances the
gravitational force affecting the particles change the sign. And the repulsion force
in a simple model of the expanded selfgraviting dust ball gives a simple and clear
explanation of the acceleration of the Universe expansion \cite{verozubA02},
\cite{verozubB02}, 

It is seems in the first sight that the accretion onto the object give rise to a too
large energy release at the surface that contradicts the low bolometric luminosity (
$\thicksim10^{36}$ $erg\ s^{-1}$ ) of Sgr A*. However, it must be taken into account
that in the gravitation theory under consideration the velocity of free falling test
particles decrease inside the Schwarzschild radius \cite{verozub91}. If we assume
that the the radius $R$ of the object with mass of $2.6\cdot M_{\bigodot}$ in the
Galactic Center is equal to $0.04$ $r_{g}$ ( which follows from the solution of the
equation of the hydrostatic  equilibrium \cite{verozub95},\cite{verkoch01},
\cite{verozub01} ), then the value of the velocity $v$ of  free falling particles at
the surface is $4\cdot10^{8}$ $cm\ s^{-2}$ . Therefore, even at the accretion rate
$\overset{\cdot}{M}=10^{-6}$ $M_{\odot}\ yr^{-1}$ the amount of the released energy
$\overset{\cdot}{M}v^{2}/2$\  is only $\thicksim10^{36}$ $erg/s$ .

\section{Atmosphere}
It is believed that the rate of gas accretion onto the supermassive object in the
Galaxy center due to the star wind from the surrounding young stars is of the order
of $10^{-7}$ $M_{\odot}\ yr^{-1}$ \cite{coker}.  Therefore, if the object has a
surface, it must have also some, mainly hydrogen, atmosphere. For $10$ $Myr$ (an
estimated lifetime of the surrounding stars) the mass $M_{atm}$ of the gas envelope
reaches $M_{\odot}.$ The height of the homogeneous atmosphere $h_{atm}=kT/ 2\ m_{p}\
g$, where $k$ is the Boltzmann constant, $T$ is the absolute temperature, $m_{p}$ is
the proton mass, $g=F/m=7.8\cdot10^{6}\ cm\  s^{-2}$. At the temperature $T=10^{7}K$
the high of the atmosphere $h_{atm}=10^{8}cm\thickapprox 0.01R.$ The density of the
atmosphere $\rho=M_{atm}/4\pi R^{2}h\thickapprox10^{2}\div10^{3}g\ cm^{-3}$. Under
the condition a hydrogen burning must begin in our time. Of course, the accretion
rate in the past it could has been many order greater than the mention above if the
stars was born in a molecular cloud at the same area where they are in our time.
(See the discussion in \cite{gensel1, ghes1}. In this case the burning began more
than $10\ Myr$ ago and is more intensive.

A relationship between the temperature and the density can be found from the
thermodynamics equilibrium equations%
\begin{equation}
\frac{1}{4\pi r^{2}\rho}\frac{dL_{r}}{dr}=\epsilon+\lambda\frac{\overset
{\cdot}{M}c^{2}}{M_{atm}}-T\frac{dS}{dt},
\end{equation}
where
\begin{equation}
L_{r}=4\pi r^{2}\frac{ac}{3\kappa\rho}\frac{d}{dr}T^{4},
\end{equation}
$\overset {\cdot}{M}$ is the rate of the mass accretion , $\epsilon$ is the rate of
the generation of the nuclear energy per the unit of mass, $\lambda$ is the portion
of the  thermalized accretion energy , $\kappa$ is the Krammers absorbtion factor,
$S$ is entropy, $a$ is the radiation constant. In the stationary case for the
homogeneous atmosphere we have%
\begin{equation}
\frac{acT^{4}}{3\kappa\rho^{2}h_{atm}^{2}}=\epsilon+\lambda\frac{\overset{\cdot}%
{M}c^{2}}{M_{atm}}.
\end{equation}
\begin{figure}
\resizebox{\hsize}{!} {\includegraphics[]{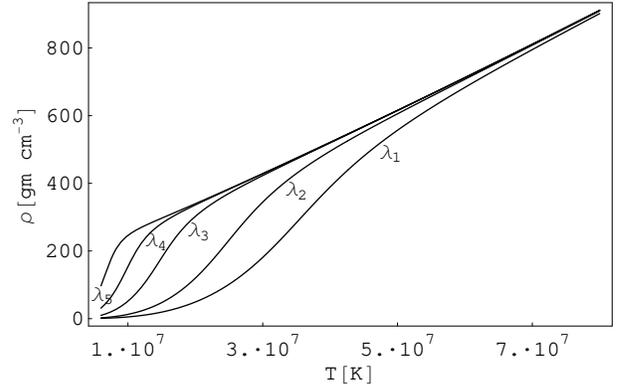}}
 \caption{Relationship between temperature and pressure of the homogeneous
atmosphere for the values of the parameter $\lambda:$ $\lambda_{1}%
=1,\lambda_{2}=1/6,\lambda_{3}=10^{-2},\lambda_{4}=10^{-3},\lambda_{5}%
=10^{-4}$.}
 \label{RelTP}
\end{figure}

Fig. \ref{RelTP} shows the relationship between $T$ and $\rho$ for several values of
the parameter $\lambda$. It follows from the figure that the hydrogen burning can be
occurs at the density $10^{2}\div10^{3}$ $gm\ cm^{-3}$.

The luminosity from the burning layer can be found by solution of the system
of the differential equations%
\begin{align}\label{AtmProfile}
\frac{dm_{r}}{dr} =4\pi r^{2},\ \frac{dP}{dr}=\rho g,\ \frac{dL_{r}}{dr}=4\pi
r^{2}\rho\epsilon ,\\ \nonumber \frac{dT}{dr}=-\frac{3\kappa\rho L_{r}}{16\pi
acr^{2}T^{3}},
\end{align}
where $P=\rho kT/2m_{p}$ is the pressure of the nogenerated gas, $m_{r}$ is the mass
of the spherical layer of the radius $r$. The boundary data  at the surface are:
$m_{r}(R)=0,$ $\rho(R)=\rho_{0},$ $T(R)=T_{0},$ $L_{r}(R)=0$. We setting
$T_{0}=10^{7}K$ and find that the luminosity $L=2.4\cdot10^{31}erg\ s^{-1}$
at $\rho_{0}=10^{2}g\ cm^{-3}$ and $L=2.4\cdot10^{33}erg\ s^{-1}$ at $\rho_{0}%
=10^{3}g\ cm^{-3}$. The effective temperature $T_{eff}$ is $2.4\cdot10^{3}K$ and
$7.8\cdot10^{3}K$ , respectively. These magnitudes may be much more if in the past
the accretion was more intensive. However, if this is not true, and if the accretion
rate is much less than $10^{-7}M_{\odot},$ the hydrogen burning presently may be
missing.
\section{Peculiarity of the accretion  }
The main peculiarity of a spherical supersonic accretion onto the supermassive
object without events horizon is the existence of the second sonic point -- in a
vicinity of the object, which does not bound with hydrodynamical effects. The
physical reason is that as the sound velocity $v_{s}$ in the infalling flow grows
together with the temperature, the gas velocity $v$ begin to decrease fast before
$r=r_{g}$ \cite{verozub91}. As a result, the equality $v=v_{s}$ take place at some
distance $r_{s}$ from the center.

In accordance with \cite{verozub91} the maximal radial velocity of a free falling
particle does not exceed $0.4\ c.$ Therefore, the Lorentz-factor is nearly $1$ and
the accretion equations are given by
\begin{align}\label{hydroeqs}
4\pi r^{2}v  &  =\overset{\cdot}{M} ,\ \ \ vv+\overset{\cdot}{n}/n=g_{fr}\\
\left(
\frac{\varepsilon}{n}\right)^{^{\prime}}-P\frac{n^{\prime}}{n}=-\frac{\Lambda}{v\
n}. \nonumber
\end{align}

In these eqs.  $n$ is the density of the particles number, $\Lambda$ is the cooling
rate (per volume unit) due to the bremsstrahlung and comptonization , $g_{fr}=F/m$
is defined by eq. (\ref{gravaccel1}). We, therefore, neglect the radiation pressure
and do not take into account the subequipartition magnetic field which may exist in
the accretion flow \cite{melia}.

The energy density is assumed to be equal to [\cite{melia}]
\begin{equation}
\varepsilon=m_{p}c^{2}n+\alpha nkT,
\end{equation}
where
\begin{align}
\alpha=3+x\left(  \frac{3K_{3}(x)+K_{1}(x)}{4K_{2}(x)}-1\right) \\ \nonumber
 +y\left(
\frac{3K_{3}(y)+K_{1}(y)}{4K_{2}(y)}-1\right)  ,
\end{align}
$x=m_{e}c^{2}/kT$, $y=m_{p}c^{2}/kT$ and $K_{j}$ ($j=2,3$) is  the modified Bessel
function.

We assume for definiteness  that $\overset{\cdot}{M}=10^{-7}\ M_{\odot}\ yr^{-1}$
and at the distance from the center $r_{0}=10^{17}cm$ the velocity
$v(r_{0})=10^{8}cm\ s^{-1}$ and the temperature $T(r_{0})=10^{5}K$ . The profile of
the temperature and velocity can be found by the solution of the eqs.
\ref{hydroeqs}. The sound velocity $v_{s}=v_{s}(r)$ is
\begin{equation}
v_{s}=\left(  \frac{\Gamma P}{\rho+P}\right)  ^{1/2},
\end{equation}
where $\rho$ is the density, the adiabatic index is
\begin{equation}
\Gamma=\frac{n}{P}\left(  \frac{\partial P}{\partial n}\right)  _{S_{p}}%
\end{equation}
and $S_{p}$ is entropy per particle. The values of $P$, $\rho$ and $\Gamma$ for a
perfect Boltzmann gas as the function of $T$ is given by the Service fitting
formulas \cite{service}. Under these conditions the gas temperature $T$ at $r=r_{s}$
is  of the order of $10^{10}K$. The postshock region lies from $r_{s}$ up to the
dense atmosphere ( $n\approx10^{15}\div10^{17}$ ) at the distance
     $z\approx R$
from the object surface where infalling gas velocity slumps. In this area releases
more part of the accretion energy. (We neglect here a difference between the height
of proton and electron stopping \cite{bildsten}. The energy release per $1\  g$ due
to protons stoping is given by \cite{alme}

\bigskip%
\begin{equation}
W=-f(x_{e})\frac{4\pi ne^{4}\lg\Delta}{m_{e}vm_{p}},
\end{equation}

where%

\begin{equation}
\Delta=\frac{3(kT)^{3/2}}{4\sqrt{\pi n}e^{3}}\ \ \  ;x_{e}=\left(  \frac
{m_{e}v^{2}}{kT}\right)  ^{1/2}%
\end{equation}
and%
\begin{equation}
f(x_{e)}=\frac{2x_{e}(2x_{e}^{2}/3-m_{e}/m_{p})/\pi^{1/2}}{1+2x_{e}(2x_{e}%
^{2}/3-m_{e}/m_{p})/\pi^{1/2}}.
\end{equation}

\begin{figure}
\resizebox{\hsize}{!} {\includegraphics[]{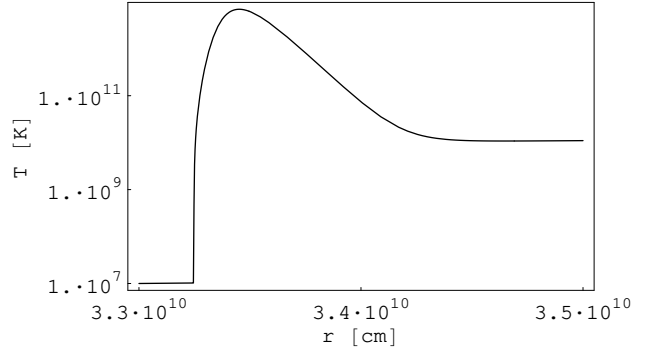}}
 \caption{A typical profile of the temperature at the deceleration area.}
 \label{JumpT}
\end{figure}

We assume that the gas density profile in this area is the sum of the atmosphere
density resulting from eqs. (\ref{AtmProfile}) and the solution $n(r)$ of the eqs.
(\ref{hydroeqs}). Let $U$ be the radiation density and $k_{2}=0.4 +6.4\cdot10^{22}
m_{p}nT^{-7/2}$ $cm^{2} g^{-1}$ is the flux mean opacity. At the given density the
profile ("jump") of the
luminosity and temperature can be found approximately from the equations%
\begin{equation}
dL_{r}/dr=4\pi r^{2}\rho W,\;\frac{1}{3\rho}\frac{dU}{dr}=k_{2}\frac{L_{r}}{4\pi
r^{2}c},
\end{equation}
and from the simple energy balance equation \cite{zeldovich} between $W$,
bremsstrahlung and comptonization processes. Fig. \ref{JumpT} shows a typical
temperature profile close the distance $z=R$ from the object surface. The following
end conditions were used: the luminosity at the distance from the surface $0.0785R$
is $10^{32}erg\ s^{-1}$, the temperature is $10^{7}K,$ the density of radiation is
$aT_{eff}^{4},$ where $T_{eff}=10^{4}K$ and $a=7.569\cdot10^{15}$.

\section{ Comptonization  } 
\begin{figure}
\resizebox{\hsize}{!} {\includegraphics[]{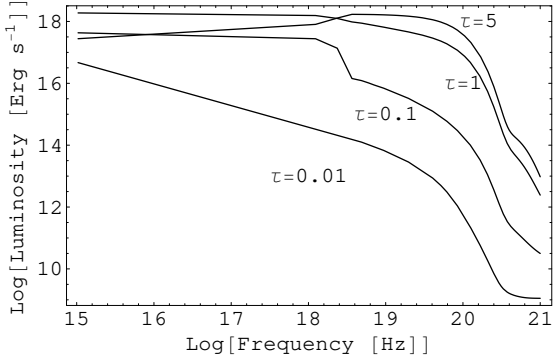}}
 \caption{The emerging spectrum of the nuclear burning after comptonization by the hot lay at
 the optical thickness $\tau=0.01,0.1,1,5$  }
  \label{ComptonSpectrum}
\end{figure}

The emerging intensity $I$ of the low atmosphere
radiation after passage through a hot spherical homogeneous lay of gas is a
convolution of the incoming Plankian intensity  $I_{0}$ and the frequency
redistribution function $\Phi (s)$ of $s=\lg(\nu/\nu_{0}),$ where $\nu_{0}$ and
$\nu$ are the frequencies of the
incoming and emerging radiation, respectively:%

\begin{equation}
I(x)=\int_{-\infty}^{\infty}\Phi(s)I_{0}(s)ds.
\end{equation}
 At the temperature of
the order of $10^{10}K$ the dimensionless parameter $\vartheta=kT/m_{e}c^{2}\sim1$
or more that that. Under the condition the function $\Phi(s)$ can be calculated as
[\cite{brinkshaw}]
\begin{equation}
\Phi(s)=\sum_{k=0}^{m}\frac{e^{-\tau}\tau^{k}}{k!}P_{k}(s),
\end{equation}
where
\begin{equation}
P_{1}(s)=\int_{\beta_{\min}}^{1}\varphi(\beta)P(s,\beta)d\beta,
\end{equation}

\begin{equation}
\beta_{\min}=\frac{e^{\shortmid s\shortmid}-1}{e^{\shortmid s\shortmid}+1}%
\end{equation}

and
\begin{align}
P(s,\beta)=\frac{3}{16\gamma^{4}\beta}\int_{\mu_{1}}^{\mu_{2}}(1-\beta\mu)^{-3}\times\\
 \left(
1+\beta\mu^{\prime})(1+\mu^{2}\mu^{\prime2}
+\frac{1}{2}(1-\mu^{2})(1-\mu^{\prime2})\right)
 d\mu, \nonumber
\end{align}%
\begin{equation}
\mu^{\prime}=\frac{e^{s}(1-\beta\mu)-1}{\beta},
\end{equation}%
\begin{equation}
\mu_{1}=\left\{
\begin{array}
[c]{c}%
-1\hspace{0.5cm}\hspace{0.5cm}\hspace{0.2cm}\hspace{0.2cm}s\leq0\\
\frac{1-e^{-s}(1+\beta)}{\beta}\hspace{0.5cm}s\geq0
\end{array}
\right.
\end{equation}%
\begin{equation}
\mu_{2}=\left\{
\begin{array}
[c]{c}%
\frac{1-e^{-s}(1-\beta)}{\beta}\hspace{0.5cm}\hspace{0.5cm}\hspace
{0.2cm}\hspace{0.2cm}s\leq0\\
1\hspace{0.5cm}s\geq0
\end{array}
\right.
\end{equation}
and $\gamma=(1-\beta^{2})^{-1/2}.$ Taking into account the scattering up to 4th
order ( $m=4)$ we find an approximate emerging spectrum of the nuclear burning with
$T_{eff}=10^{4}K$ after  comptonization for several values of the optical thickness
$\tau$  of a homogeneous  hot layer. The results are plotted in fig.
\ref{ComptonSpectrum}

It follows from the figure for small magnitudes of $\tau$ that the spectrum  can
contribute to the observed Sgr A* radiation \cite{baganoff}, \cite{porquet},
\cite{genzel3}  in IR and X regions.

It should be noted that the
 timescale $\Delta t$ of a variable process are related to the size $\Delta r$
 of its region as $ \Delta r\leq c\ \Delta t/(1+z_{g})$, where $1+z_{g}=1/\sqrt{C(r)}$
 is
 gravitational redshift. (The function $C(r)$ is defined by eqs. \ref{ABC}. Therefore, the variations
 in the radiation intensity $\sim 600\ s$ happen at the
distance $z\leqslant 1.7 R$ from the surface. For this reason IR and X flares
\cite{baganoff}, \cite{porquet}, \cite{genzel3} can be interpreted as processes near
the surface of the objects.

\section{Transfer equation and synchrotron radiation of Sgr A* }

Some authors \cite{mitra} ,\cite{robertson} have found evidence for  existence of an
intrinsic magnetic field of Black Holes candidates that is incompatible with the
events horizon. Here we calculate the contribution in the synchrotron radiation
spectrum from the intrinsic magnetic field of the form
\begin{equation}\label{ProperB}
B=B_{0}  \left( \frac{R}{r}\right)^{3},
\end{equation}
where $B_{0}=2\cdot 10^{5}\  Gs$.
 For study of the radiation from the  surface vicinity of the supermassive object
 the influence of
gravitation on the frequency of the photon and its motion must be taken into
account. A relativistic transfer equation can be used for this purpose. It is a
relativistic Boltzmann  equations for photon gas \cite{lindkvist},
\cite{shmidt-burg}, \cite{zane}. We assume that in the spherically symmetric field
the distribution function $\mathcal{F}$ depends on the radial distance from the
center $r$ , the frequency $\nu$ and the photon direction which can be defined by
the cosine of the horizontal angle -- $\mu$. (The spherical coordinate system is
used). The relativistic Boltzmann equation in Minkowski space - time is of the form

\begin{equation}
\frac{d\mathcal{F}}{d\eta}=St(\mathcal{F}), \label{Bolz1}%
\end{equation}
where $d\mathcal{F}/d\eta$ is the   derivative along the 4-trajectory $ x^{\alpha}=
x^{\alpha}(s)$
 in space-time with the metric differential form $d\eta$ and the right-hand member
is the collisions integral. By using for photons  $x^{0}=c\ t$  as a parameter along
4-trajectory we arrive at the transfer equation in the form
\begin{equation}
\left(  \frac{\partial}{\partial dx^{0}}+\mathbf{n\nabla+}\frac{d\mu
}{dx^{0}}\frac{\partial}{\partial\mu}+\frac{d\nu}{dx^{0}}\frac{\partial
}{\partial\nu}\right)  \mathcal{F}=St(\mathcal{F}), \label{TransferEq0}%
\end{equation}
where the magnitudes $d\mu/dx^{0}$ and $d\nu/dx^{0}$ must be found from our
equations of the photon motion in the gravitation field \cite{verozub95} . The
collisions integral is given by
\begin{equation}
\chi\text{ }(S/\beta-\mathcal{F}),
\end{equation}
where $\chi$ is the absorption coefficient,  $S$ $=\eta/\chi$ is the source
function, $\eta$ is the emissivity and $\beta$ $=h^{4}\nu^{3}/c^{2}$. In this paper
we do not take into account light diffusion. The intensity $I$ of the radiation is
related to $\mathcal{F}$ as $I=\beta F$.

To solve the transfer equation we  use the characteristic method \cite{shmidt-burg},
\cite{zane}. Since photon's trajectories are  characteristics of the partial
differential equation (\ref{TransferEq0}), the equations are reduced to ordinary
differential equations along these trajectories. In
our case this equation is%

\begin{equation}
\frac{d\mathcal{F}}{dr}=\frac{c}{v_{ph}}\left(\frac{S}{\beta}
-\mathcal{F}\right), \label{TransferEq1}%
\end{equation}
where $v_{ph}$ is the radial photon velocity.

According to [\cite{verozub91}] in the spherically - symmetric field
\begin{equation}
v_{ph}=c\sqrt{\frac{C}{A}\left(  1-C\frac{b^{2}}{f^{2}}\right)  },
\end{equation}
where $b$ is the impact parameter of photon.

For numerical estimates we assume that a relativistic Maxwellian electron
distribution take place and set emission coefficient \cite{wardzinski}

\begin{equation}\label{}
    j_{\nu}=\frac{2^{1/6} \pi^{3/2} e^{2} n_{e} \nu)}{3^{5/6} c K_{2}(1/\Theta)
    \Upsilon^{1/6}} exp[-(9 \Upsilon/2)^{1/3}],
\end{equation}
where $\Upsilon=\nu/\nu_{c} \Theta^{2}$ and $\Theta=k T/m c^{2}$, $n_{e}$ is the
electron density number, $K_{2}$ is the modified Bessel function of the second kind.

Let $\mathcal{F}_{\nu}(b)$ be the solution  of the differential equation
(\ref{TransferEq1}) for a given $b$ at $r\rightarrow \infty$. Then, for a distant
observer the luminosity at the frequency $\nu$ is
given by%

\begin{equation}
L_{\nu}=8\pi^{2}\int_{0}^{\infty}\beta\mathcal{F}_{\nu}(b)bdb.
\end{equation}

\begin{figure}
\resizebox{\hsize}{!} {\includegraphics[]{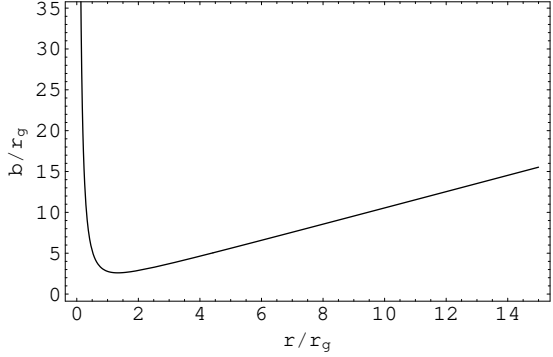}}
 \caption{The function $b(r)$}
 \label{bPlot}
\end{figure}

For a correct solution of eq.(\ref{TransferEq1}) it is essential that there are
three types of photons trajectories in the spherically-symmetric gravitation field
in view the used gravitation equations . It can be seen from fig. \ref{bPlot}. It
shows the geometrical locus where the radial photon velocities are equal to zero
which is given by the equation
\begin{equation}
b=\frac{f}{\sqrt{C}}. \label{FotonLocus}%
\end{equation}

The minimal value of $b$ is $b_{cr}=3\sqrt{3}r_{g}/2$. It occurs at the distance
from the center $r_{cr}=\sqrt[3]{19}r_{g}/2.$ The photons whose impact parameter
$b<b_{cr}$ can freely move from the object surface to infinity. The photons with $b
< b_{cr}$ cannot go away from the surface to infinity. At last,  photons with
$b>b_{cr}$ can move to infinity only if their trajectories begin at distances $r>$
$r_{cr}$. For a given $b$ the corresponding magnitude of $r$ can be found from
eq.(\ref{FotonLocus})

The differential equation for photon trajectories with $b<$ $b_{cr}$ were integrated
at the edge condition $\mathcal{F}(R)=0$. The distribution function $\mathcal{F}$ at
the points of the curve in fig. \ref{bPlot} at $r>r_{cr}$ was found by solution of
differential equation (\ref{TransferEq1}) by used the end condition $\mathcal{F}=0$
at infinity. Similarly to [\cite{zane}] these magnitudes were used as end conditions
for the solution of the eq. (\ref{TransferEq1}) to find the emergent radiation.

\begin{figure}
\resizebox{\hsize}{!} {\includegraphics[]{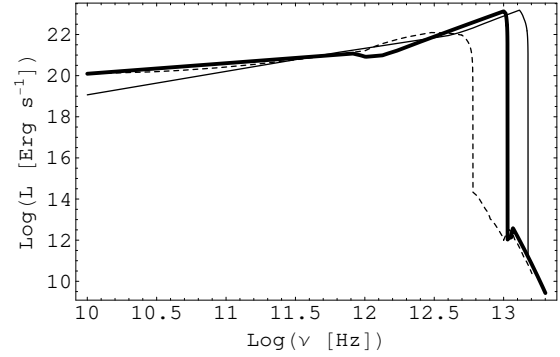}}
 \caption{The spectrum of the synchrotron radiation caused by the proper
magnetic field of the object.}
 \label{spectrumS}
\end{figure}

Fig.\ref{spectrumS} shows the spectrum of the synchrotron radiation in the band
$10^{10}\div2\cdot10^{13}\ Hz$ for three cases:

-- for the intrinsic magnetic field (\ref{ProperB}) of the object (the thin line),

-- for the subequipartition magnetic field which may exist in the accretion flow
\cite{melia} $B_{ext}=(\dot{M}v/4\pi r^{2})^{1/2}$ ($v$ is the gas velocity at the
accretion) (dash line).

--  for the sum of the above magnetic fields (thick curve).

It follows from the figure that  existence of the intrinsic magnetic field may give
a good fitting to the observation data.
\section{Conclusion}
The assignment of  nature of compact objects in the galactic  centers is one of
basic problems of fundamental physics and astrophysics. The results received above,
certainly, yet do not allow to draw the certain conclusions. However they show that
the opportunity investigated here does not contradicts the observant data, and,
therefore, demands the further study.

\end{document}